\documentclass[aps,pra,superscriptaddress,twocolumn,longbibliography]{revtex4-1}
\usepackage{mathrsfs}
\usepackage{amsfonts}
\usepackage{amssymb}
\usepackage{amsmath}
\usepackage{graphicx}

\graphicspath{{figures/}{./}} 
\usepackage{sidecap}
\usepackage{color}
\usepackage{xcolor}
\usepackage[colorlinks=true, urlcolor=blue, citecolor=blue,linkcolor=blue,citebordercolor={1 0 0},linkbordercolor={0 0 1}]{hyperref}
\usepackage{subeqnarray}
\usepackage{verbatim}
\usepackage{euscript} 

\usepackage{hyperref}

\usepackage{bibunits}

\usepackage{array}
\usepackage{multirow}

\newcommand{\ket}[1]{\vert #1 \rangle}
\newcommand{\bra}[1]{\langle #1 \vert}
\newcommand{\ketbra}[2]{\vert #1 \rangle \langle #2 \vert}
\newcommand{\braket}[2]{\langle #1 \vert #2 \rangle}

\renewcommand{\eqref}[1]{Eq.~(\ref{#1})} 
\newcommand{\figref}[1]{Fig.~\ref{#1}} 
\newcommand{\secref}[1]{Sec.~\ref{#1}} 

\newcommand{\intelcali}{Intel Labs, Santa Clara, California 95054, USA}
\newcommand{\utchem}{Department  of  Chemistry,  University  of  Toronto,  Toronto,  Ontario  M5G 1Z8,  Canada}
\newcommand{\utcomp}{Department  of  Computer Science,  University  of  Toronto,  Toronto,  Ontario  M5S 2E4,  Canada}
\newcommand{\vectorinst}{Vector  Institute  for  Artificial  Intelligence,  Toronto,  Ontario  M5S  1M1,  Canada}
\newcommand{\cifar}{Canadian  Institute  for  Advanced  Research,  Toronto,  Ontario  M5G  1Z8,  Canada}

\begin{document}
\title{Quantum computer-aided design: digital quantum simulation of quantum processors}

\author{Thi Ha Kyaw}
\thanks{These authors contributed equally to this work.\\ \urlstyle{same} \url{thihakyaw@cs.toronto.edu} \, \url{tim\_menke@g.harvard.edu}}
\affiliation{\utcomp}
\affiliation{\utchem}
\author{Tim Menke}
\thanks{These authors contributed equally to this work.\\ \urlstyle{same} \url{thihakyaw@cs.toronto.edu} \, \url{tim\_menke@g.harvard.edu}}
\affiliation{Department of Physics, Harvard University, Cambridge, MA 02138, USA}
\affiliation{Research Laboratory of Electronics, Massachusetts Institute of Technology, Cambridge, MA 02139, USA}
\affiliation{Department of Physics, Massachusetts Institute of Technology, Cambridge, MA 02139, USA}
\author{Sukin Sim}
\affiliation{Department of Chemistry and Chemical Biology, Harvard University, Cambridge, MA 02138, USA}
\author{Abhinav Anand}
\affiliation{\utchem}
\author{Nicolas P. D. Sawaya}
\affiliation{\intelcali}
\author{William D. Oliver}
\affiliation{Research Laboratory of Electronics, Massachusetts Institute of Technology, Cambridge, MA 02139, USA}
\affiliation{Department of Electrical Engineering and Computer Science, Massachusetts Institute of Technology, Cambridge, MA 02139, USA}
\author{Gian Giacomo Guerreschi}
\affiliation{\intelcali}
\author{Al\'an Aspuru-Guzik}
\email{alan@aspuru.com}
\affiliation{\utcomp}
\affiliation{\utchem}
\affiliation{\vectorinst}
\affiliation{\cifar}

\date{\today}

\begin{abstract}
With the increasing size of quantum processors, sub-modules that constitute the processor hardware will become too large to accurately simulate on a classical computer. 
Therefore, one would soon have to fabricate and test each new design primitive and parameter choice in time-consuming coordination between design, fabrication, and experimental validation. 
Here we show how one can design and test the performance of next-generation quantum hardware – by using existing quantum computers.
Focusing on superconducting transmon processors as a prominent hardware platform, we compute the static and dynamic properties of individual and coupled transmons. 
We show how the energy spectra of transmons can be obtained by variational hybrid quantum-classical algorithms that are well-suited for near-term noisy quantum computers.
In addition, single- and two-qubit gate simulations are demonstrated via Suzuki-Trotter decomposition. 
Our methods pave a promising way towards designing candidate quantum processors when the demands of calculating sub-module properties exceed the capabilities of classical computing resources.    
\end{abstract}

\maketitle


\section{Introduction}
\label{sec:introduction}

In the semiconductor industry, next-generation microprocessors are developed using existing computers and a variety of software tools for hardware description and simulation \cite{gielen2000computer,cox1992code,tuma2009circuit,austin2002simplescalar,larson2001mase}.
During this design process, it has been established that a \textit{less} powerful processor or a collection of such processors is capable of aiding the design of a novel and \textit{more} powerful processor.
The existing processor does not need to fully simulate the next-generation one; it suffices to use the current processor to understand and validate design primitives of the future processor's sub-modules. 
We refer to a sub-module as tightly linked circuitry that shares a relatively large amount of wiring, forms a small building block on its own, and is often repeated throughout the processor.
Similarly, we observe that quantum processor sub-module design and simulation has been performed on classical computers \cite{burkard2004multilevel, vool2017introduction, krantz2019quantum, menke2019automated, li2020towards}.
The resource requirements for such classical simulations scale exponentially with the number of simulated qubits and with the number of energy levels included per qubit. 
For example, superconducting transmon qubits have low anharmonicity between the energy levels, and inclusion of additional states beyond the computational states is important in order to determine the device characteristics and design high-fidelity gate pulses \cite{krantz2019quantum}.
Exact diagonalization of the transmon circuit Hamiltonian shows that a numerically accurate simulation requires about $16$ basis states.
Other states can be neglected because they do not contribute significantly to the relevant low energy eigenstates.
Therefore, the truncated Hilbert space of a processor with $M$ transmons has dimension $16^M$.
As a result, classical simulations of quantum hardware are limited to a small number of qubits, even when employing the most powerful supercomputers.
With the recent growth of quantum processor sizes, hardware simulations are already becoming infeasible.
We contend that this challenge can be addressed by simulating the quantum processor hardware on a quantum computer. 
Similarly, Ref. \cite{iyer2018small} suggests to use a quantum computer simulation rather than classical one to optimize a fault-tolerant quantum algorithm performance with a targeted logical fault rate, since the noise parameters in open quantum system simulations of the entire fault-tolerant protocol grow exponentially with the number of qubits.

\begin{figure*}[htb]
\centering
\includegraphics[scale=1.0]{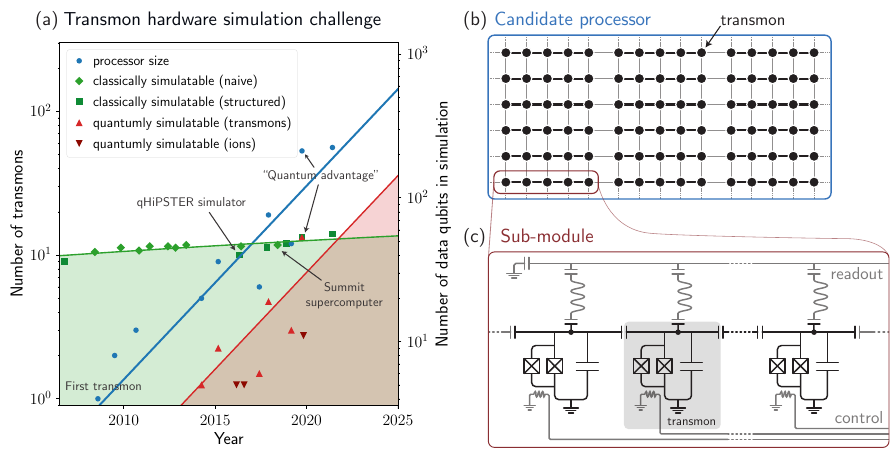}
\caption{
The challenge of simulating transmon quantum processors.
(a) The historical numbers of transmons in a superconducting processor suggest an exponential growth in processor size, while classical supercomputer simulations will be limited to about ten transmons in the forseeable future.
Quantum computers may soon be able to simulate larger transmon processors than their classical counterparts.
The blue dots show the size of a selection of published transmon processors over time (Refs. \cite{houck2008controlling,dicarlo2009demonstration,dicarlo2010preparation,barends2014superconducting,kelly2015state,kandala2017hardware,otterbach2017unsupervised,gong2019genuine,google_quantum_supremacy,wu2021strong} chronologically).
The green diamonds show the maximum number of transmons that the best supercomputer at the given point in time could simulate based na\"ively on its memory size, using a Hilbert state truncation of 16 states per transmon \cite{top500}.
Another estimate for the classically simulatable processor size takes the maximum number of data qubits for which a quantum algorithm has been simulated on a supercomputer and scales it to the number of transmons that these qubits could encode (green squares, Refs. \cite{raedt2007massively,smelyanskiy2016qhipster,haener2017petabyte,raedt2019massively},\cite{google_quantum_supremacy,wu2021strong} chronologically).
An encoding scheme with four data qubits per transmons is assumed.
The red triangles show the number of transmons that could be simulated on an experimentally demonstrated quantum computer that is made up of transmons themselves (triangles up, Refs. \cite{barends2014superconducting,kelly2015state,kandala2017hardware,otterbach2017unsupervised,gong2019genuine,google_quantum_supremacy,wu2021strong} chronologically)
or of ions (triangles down, Refs. \cite{monz2016realization,debnath2016demonstration,wright2019benchmarking} chronologically).
All solid lines are to guide the eyes along the general trend of the transmon processor size (blue), classically simulatable processor size (green), and quantumly simulatable processor size (red).
Note that the cost of simulation is a constant offset in this log-linear plot and grows as fast as the available quantum processors.
(b) Design proposals for new processors typically consist of tiled sub-modules (red box) that share common readout circuitry and that can be perceived as the unit cells of the processor.
(c) This work considers the case in which the sub-modules are chains of capacitively coupled transmons, for which the circuit diagram including readout and flux control circuitry is shown here.
}
\label{fig:intro}
\end{figure*}

By ``quantum computer-aided design” we refer to the use of quantum computers to generate the energy spectrum of a quantum device with high precision as well as to simulate its unitary dynamics in order to facilitate the design of quantum processors.
Taking transmon processors as an example (see \figref{fig:intro}(b-c)), we develop explicit quantum algorithms for hardware simulation, provide resource counts, and give a quantitative estimate for when quantum hardware simulation capabilities will surpass those of classical simulations.
As shown in \figref{fig:intro}(a), the number of transmons in experimentally demonstrated processors grows at an exponential pace.
A selection of processors that are capable of shallow algorithms or processor-wide entanglement is visualized, from the first demonstration of the transmon in 2008 \cite{schreier2008suppressing} to the ``beyond-classical quantum advantage" experiments \cite{google_quantum_supremacy,wu2021strong}.
We contrast the growth of the transmon processor size with the number of transmons that can be simulated on a classical computer.
The memory size of the most powerful supercomputers is sufficient to store the relevant state space of about ten coupled transmons.
As each additional transmon requires a 16-fold increase in memory, the effectiveness in simulation capacity of supercomputers has been slow over the past decade.
Although the structure of a quantum computation problem can often be exploited to reduce the classical simulation resource requirements \cite{di2019efficient}, the required memory will still grow exponentially in transmon number if one seeks to study a fully interacting processor.
These trends illustrate the challenge of validating new quantum processor designs computationally before passing them on to the cost- and time-intensive fabrication and experimental verification stages.

The concepts applied in this work are generalizable to other transmon-based quantum hardware as well as to simulating the realistic multi-state nature of trapped ion \cite{bruzewicz2019trapped}, neutral atom \cite{briegel2000quantum}, and photonic systems \cite{slussarenko2019photonic}.
In a separate and similar line of work, some of us target quantum computer aided design of quantum optical hardware, translating the corresponding unitary operators to digital quantum circuits \cite{jakob}.

\section{Methods}
\label{sec:methods}

Digital quantum simulation is one of the main applications of quantum computing \cite{feynman1982simulating, lloyd1998universal}.
\figref{fig:intro}(a) shows how many transmons recent superconducting and ion based digital quantum computers could simulate, based on their number of qubits.
We divide the number of qubits in these processors by four, since we need four data qubits for a numerically accurate simulation of a single transmon with 16 basis states.
Throughout this work, we use the term ``physical qubit” to refer to the hardware (transmon) qubit that is being simulated. 
To perform the hardware simulation,  we employ architecture-independent ``data qubits” in near- and intermediate-term quantum computers to digitally simulate one or more physical qubits.
The quantum simulation algorithms that are carried out with the data qubits are applicable across quantum computing  platforms,  such as trapped ion or superconducting quantum processors.
The quantumly simulatable transmon processor size naturally follows the demonstrated experimental transmon processor size, as new digital processors can again be employed to simulate smaller processor hardware.
Therefore, the quantum simulation capacity for transmons grows much more rapidly than the classical one.

\subsection{Encoding of superconducting circuit Hamiltonians into multi-qubit operators}
\label{subsec:encoding}

We choose a quantum processor sub-module that is a chain of $M$ capacitively coupled transmons as shown in \figref{fig:intro}(c).
Each transmon \cite{koch2007charge,schreier2008suppressing} is associated with a circuit node that is indicated by a circular dot in the diagram.
The sub-module choice is motivated by a conceivable processor in which all transmons in a chain are coupled to a common readout resonator, and the separate chains that constitute the processor have weak intra-chain coupling.
Within a chain, the presence of the resonator adds to the capacitative cross-talk between the qubits and can lead to unwanted or uncontrolled coupling effects \cite{yanay2019realizing}.
The inter-chain couplings would be weaker and better controlled by comparison, so a focus on the isolated chain as a sub-module instance is warranted.

The transmon chain is described by the following Hamiltonian:
\begin{eqnarray}
\label{eq:chain}
    \hat{H}_{M\text{-transmon}} = && \, 2e^2 \sum_{i,j=1}^{M} \left(\mathbf{C}^{-1}\right)_{ij} \, \hat{N}_i \hat{N}_j \nonumber\\
     && - 2 \sum_{i=1}^M E_{\text{J},i} \left| \cos\left( 2\pi f_i \right) \right| \cos\hat{\varphi_i}.
\end{eqnarray}
Here, $e$ is the electron charge.
The normalized external flux $f_i = \Phi_{\text{ext},i} / \Phi_0$ is derived from the external magnetic flux $\Phi_{\text{ext},i}$ that penetrates the loop formed by the two Josephson junctions of each transmon.
The Josephson energy of the two junctions is equal and given by $E_{\text{J},i}$.
The magnetic flux quantum $\Phi_0$ is a fundamental constant that describes the smallest amount of flux that a superconducting loop can sustain.
The Hamiltonian is written in terms of the canonically conjugate phase operators $\hat{\varphi}_i$ and Cooper pair number operators $\hat{N}_i$ for the transmon with index $i \in \{1, ..., M\}$ (Appendix \ref{app:sec:cooper_pair_basis}).
They fulfill the commutation relation $[\hat{\varphi}_i, \hat{N}_j]= i\delta_{ij}$.
These operators are related to the flux operator $\hat{\Phi}_i= \Phi_0 \hat{\varphi}_i / 2\pi$, which denotes the flux across the Josephson junction of transmon $i$, and the charge operator $\hat{Q}_i=2e\hat{N}_i$ that describes the sum of charges stored on all capacitors connected to the transmon \cite{vool2017introduction}.
The capacitance matrix $\mathbf{C}$ is determined from the Legendre transformation of the circuit Lagrangian.
It contains the sum of all capacitances connected to a node on the diagonal and minus one times the capacitance between two nodes on the off-diagonal.
The interaction of transmons in the chain is determined by the inverse capacitance matrix $\mathbf{C}^{-1}$, which couples their charge degrees of freedom.
Although $\mathbf{C}$ is tridiagonal for a chain, its inverse is not.
Instead, for realistic parameter settings, it is a full matrix with exponentially decaying elements away from the diagonal.
Therefore, nearest-neighbor capacitances lead to non-nearest-neighbor interactions.
In addition, spurious capacitances such as those arising from a readout resonator can lead to enhanced coupling between non-nearest-neighbor sites \cite{yanay2019realizing}.
To investigate such effects prior to chip fabrication, simulation of multiple coupled transmons is necessary.
This is one of the motivations for this work on quantum simulation techniques for sub-module sizes that cannot be precisely simulated classically.

To numerically demonstrate quantum simulation of transmon sub-modules, we first restrict ourselves to one-transmon (see \figref{fig:single_transmon_VQD}(a)) and two-transmon (see \figref{fig:two_transmon_VQD}(a)) circuits.
This is sufficient to showcase quantum computation of static transmon properties as well as dynamical simulation of single-qubit and entangling gates.
The single and coupled transmon Hamiltonians can be described as a single and pair of non-linear oscillators, respectively, which is a direct result of \eqref{eq:chain}.
However, only the low-lying energy states are relevant when the transmon qubit is operated at millikelvin temperatures, given the typical excitation frequencies. 
By truncating the Hilbert space to a finite dimension, we can map the system Hamiltonians to a linear combination of products of Pauli operators.
In Appendix \ref{app:sec:cooper_pair_basis}, we discuss how to represent the multi-level operators of $\hat{H}_\text{1-transmon}$ and $\hat{H}_\text{2-transmon}$ in the charge basis first, and then transform them into a linear combination of Pauli strings in Appendix \ref{app:sec:encoding}.
The Hamiltonian then assumes the form $\hat{H} = \sum_{i} \hat{h}_i$, where each term $\hat{h}_i$ is a tensor product of multiple Pauli and identity operators, scaled by a prefactor.
While several possible mappings are available for the Pauli string transformation, the Gray encoding \cite{nicolas2019} is resource-efficient for the types of operators we consider and is used throughout this work.

\subsection{Variational quantum algorithm for multi-level systems}
\label{subsec:vqd}
With the transmon Hamiltonian rephrased in terms of qubit operators, one can estimate its eigenenergies with high accuracy by employing the VQE algorithm \cite{peruzzo2014variational, mcclean2016theory}, a hybrid quantum-classical algorithm designed to compute properties of Hermitian operators using current and near-term quantum devices.
Early implementations of VQE have focused on estimating the ground state energy of quantum systems: The quantum device is used to prepare the candidate state by executing a parameterized quantum circuit $\hat{U}_{\boldsymbol{\theta}}$ whose parameter values $\boldsymbol\theta$ are updated using classical optimizers.
The classical optimization is guided by minimizing the energy expectation value of the candidate state, usually estimated by averaging measurements from many runs of the same circuit.
The objective of VQE can be expressed as
\begin{equation}
\underset{\boldsymbol\theta}{\text{min}} \ \bra{\psi_{\boldsymbol\theta}} \hat{H} \ket{\psi_{\boldsymbol\theta}},
\end{equation}
where $\hat{H}$ corresponds to the system Hamiltonian \footnote{For all reported VQE and VQD simulations, we employed the resource-efficient Gray encoding method for mapping the system Hamiltonian onto Pauli strings.}, $\boldsymbol\theta$ corresponds to the vector of circuit parameters, and $\ket{\psi_{\boldsymbol\theta}}=\hat{U}_{\boldsymbol{\theta}}\ket{\phi_0}$ is the resulting quantum state parameterized by $\boldsymbol\theta$, where $\ket{\phi_0}$ is a reference state. 
Instances of the VQE algorithm have been demonstrated using various quantum computing architectures including superconducting circuits \cite{O_Malley2016, kandala2017hardware}, trapped ions \cite{Shen2015, Hempel2018, Nam2019}, and photonic chips \cite{peruzzo2014variational}. 
In recent years, the VQE algorithm has been generalized to estimate low-lying excited states in addition to the ground state \cite{Colless2018, Higgott2019}. In the latter work, the original objective function in VQE is extended such that the $k$-th excited state can be systematically estimated. 
This is crucial in determining higher energy levels of superconducting transmon systems, which are needed to generate high fidelity one- and two-qubit gate operations.
Similar to the VQE cost function, the objective for the $k$-th excited state of the so-called VQD algorithm \cite{Higgott2019} can be written as

\begin{equation}
\underset{\boldsymbol\theta_k}{\text{min}} \left(\bra{\psi_{\boldsymbol\theta_k}} \hat{H} \ket{\psi_{\boldsymbol\theta_k}} + \sum_{i=0}^{k-1} \beta_i |\braket{\psi_{\boldsymbol\theta_k}}{\psi_{\boldsymbol\theta_i}}|^2 \right) ,
\end{equation}

\noindent where $\boldsymbol\theta_k$ corresponds to the parameter setting that approximates the $k$-th excited state energy, and $\beta_i$ are constants that penalize nonzero overlaps with previous eigenstate estimates.
Given a large enough $\beta_i$, i.e. $\beta_i > E_k - E_i$ and a sufficiently flexible and efficient ansatz, the VQD objective can converge to $E_k$.
A self-correcting procedure is suggested for choosing an appropriate value for $\beta_i$. This involves choosing a test parameter $\gamma := \beta_i + E_i$, such that if $\gamma$ is too small, i.e. $\gamma - E_i = \beta_i \leq E_k - E_i$, the VQD algorithm outputs a minimum at $\gamma$. In such circumstances, we follow the recommended heuristic rule to double the value of $\gamma$.
All VQE and VQD wavefunction simulations are implemented using OpenFermion \cite{openfermion}, Forest \cite{quil}, Tequila \cite{kottmann2021tequila} and Qulacs \cite{suzuki2020qulacs}.

\subsection{Suzuki-Trotter simulation of quantum gates}
\label{subsec:trotter}
In addition to the static energy spectrum, simulating dynamical gate operations is an important aspect in the design of quantum hardware.
The relevant metric for a hardware gate is the gate fidelity, which is a quantitative measure of how well a control pulse transforms an initial state $\ket{\psi_0}$ into a desired state $\ket{\psi_\text{des}} = \hat{U}_\text{des} \ket{\psi_0}$, where $\hat{U}_\text{des}$ is the desired gate unitary.
Since hardware gates exhibit a number of imperfections such as inducing leakage to higher excited states, state evolution under the control pulse will not precisely implement the desired unitary $\hat{U}_\text{des}$.
The exact time evolution of the initial state $\ket{\psi_0}$ under the control pulse is written as $\ket{\psi_\text{ex}(t)}$.
The gate fidelity is then defined as $F=|\braket{\psi_\text{ex}(t_\text{g})}{\psi_\text{des}}|^2$, where the gate time $t_\text{g}$ is optimized to maximize the gate fidelity for a given control pulse.
An estimate of the average gate fidelity $\bar{F}$ over all states in the qubit Hilbert space is obtained by averaging the gate fidelity over sufficiently many Haar random initial state samples.
For the hardware gates studied in this work, we simulate the numerically exact state evolution in QuTiP \cite{johansson2012qutip} in order to estimate the gate fidelities and provide an exact solution benchmark.

We investigate the use of Suzuki-Trotter evolution to simulate hardware gates on a digital quantum computer.
With the first order Suzuki-Trotter approximation, the exact time evolution $\hat{U}_\text{ex}$ under the control pulse becomes \cite{wiebe2010higher}
\begin{eqnarray}
\label{eq:Trotterized_unitary}
    \hat{U}_\text{ex}(t) &&= \exp\left( -i\hat{H} t / \hbar \right)\nonumber \\
     &&\approx \left[ \prod_{i=1}^{N} \exp\left( -i \hat{h}_i t / \hbar K \right) \right]^K + \mathcal{O}(t^2/K) \\
     &&= \hat{U}_\text{trot}(t),\nonumber
\end{eqnarray}
where $K$ is the number of Trotter steps, $\hbar$ is the reduced Planck constant, and each exponential of a weighted Pauli string $\hat{h}_i$ is translated into a gate sequence on the digital quantum computer.
The state evolution $\ket{\psi_\text{trot}(t)} = \hat{U}_\text{trot}(t) \ket{\psi_0}$ is simulated numerically using the Intel Quantum Simulator \cite{guerreschi2020intel, smelyanskiy2016qhipster}.
In order to assess the resource requirements for Suzuki-Trotter simulations on digital quantum computers, it needs to be determined how the simulation error depends on the Trotter step size $\Delta t = \frac{t}{K}$.
While the theoretical error $\mathcal{O}(t^2/K)$ of the approximate time evolution operator provides an upper bound that is linear in $\Delta t$, numerical simulations can often tighten this bound \cite{kivlichan2019improved}.
We define the Trotter error of a specific simulation as $\mathcal{E} = 1 - |\braket{\psi_\text{trot}(t_\text{g})}{\psi_\text{ex}(t_\text{g})}|^2$ and again approximate its average $\bar{\mathcal{E}}$ over the qubit Hilbert space by Haar random sampling of initial states.

The formulation of \eqref{eq:Trotterized_unitary} assumes a time-independent Hamiltonian.
In the case of a time-dependent Hamiltonian $\hat{H}(t)$ such as for the single-qubit pulse, the Trotter error also depends on the time derivative of the Hamiltonian \cite{low2018hamiltonian}.
As the time dependence in the single-qubit case is slow, we do not expect a large added error and refer to Refs.~\cite{wiebe2010higher,poulin2011quantum,low2018hamiltonian} for further information on time-dependent Suzuki-Trotter algorithms and error budgets.


\section{Results}

This section is divided into single- and two-transmon design as well as a scaling analysis for larger processor sizes.
We discuss initial state preparation via the VQE and VQD algorithms and Trotterized evolution to carry out dynamical quantum simulations.
Based on the energy spectrum and eigenstates, we can analyze whether or not the quantum device being designed is robust against some external controllable or uncontrolled parameters \cite{yan2016flux}. 
In addition, dynamical simulations aid in the design of desired quantum gates.

\begin{figure}[htb]
\centering
\includegraphics[scale=0.4]{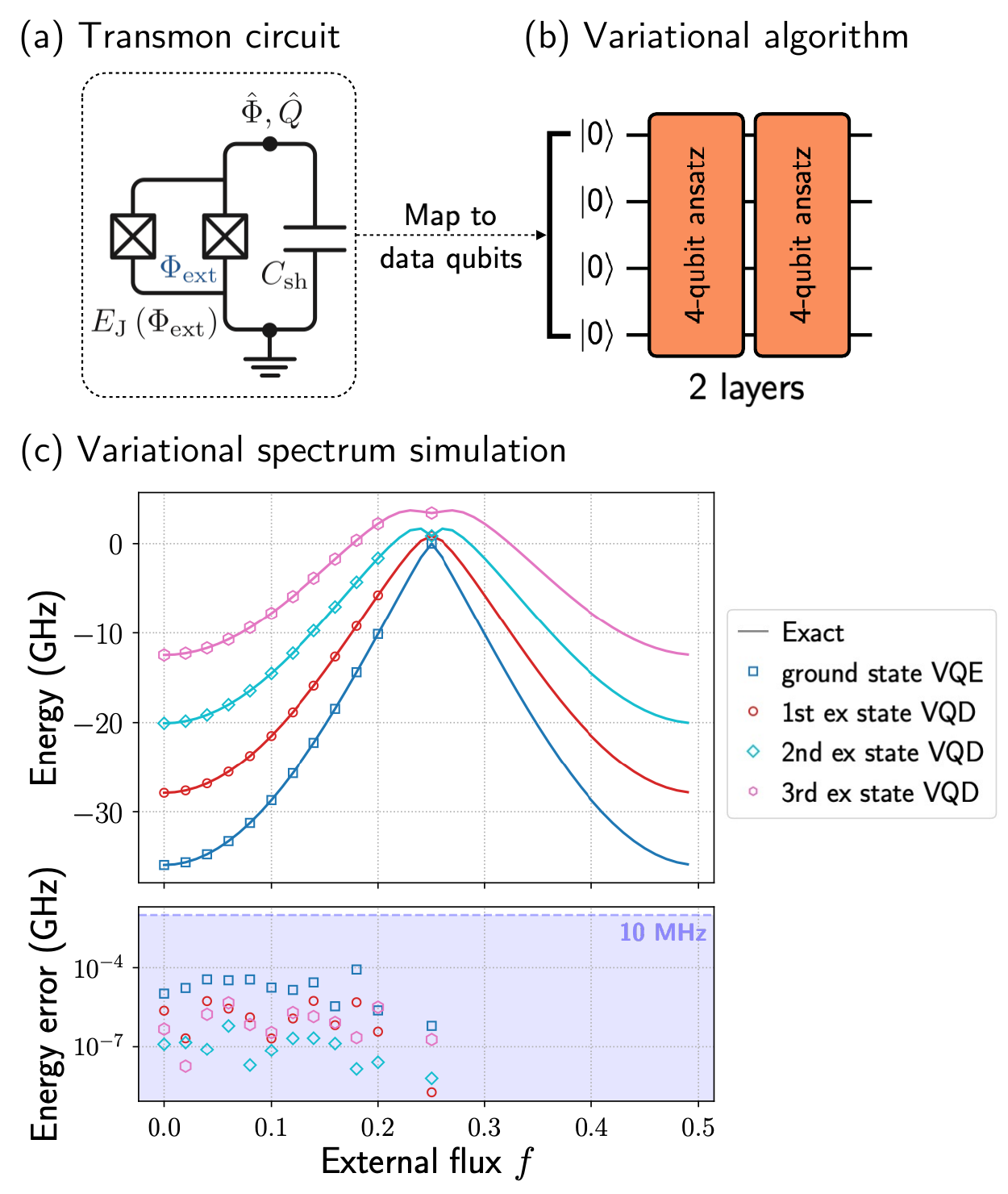}
\caption{(a) Circuit diagram of a transmon qubit with tunable external magnetic flux $\Phi_{\textrm{ext}}$.
(b) After mapping the transmon Hamiltonian to four data qubits, a variational algorithm is applied to find the ground state and first few excited states of the circuit.
(c) The four lowest eigenenergies are computed against the normalized external magnetic flux $f=\Phi_{\textrm{ext}}/\Phi_0$, where $\Phi_0$ is the magnetic flux quantum. 
The solid lines correspond to the exact diagonalization result, for which the basis of the circuit Hamiltonian is truncated to $d=16$ states.
Colored markers show the quantum simulation estimates using the VQE and VQD algorithms.
Energy errors of the quantum algorithms are also determined:
Each dot represents the error incurred after completion of the algorithms.
The shaded region indicates an error threshold of 10\,MHz and below.}
\label{fig:single_transmon_VQD}
\end{figure}

\subsection{Single-transmon quantum simulation}
\label{subsec:single-transmon}

We consider a flux tunable transmon consisting of a two-junction SQUID loop with Josephson energy $E_\text{J}/h = 20\,\text{GHz}$ per junction, where $h$ is the Planck constant, and capacitance $C_\text{sh}+C_\textrm{J} = 91\,\text{fF}$.
The ratio of the Josephson and capacitive energy is $E_\text{J}/E_\text{C} =94.0$, where $E_\text{C} = e^2 /2(C_{\textrm{sh}}+ C_{\textrm{J}})$.
A circuit diagram of the transmon is shown in \figref{fig:single_transmon_VQD}(a).
The effective Josephson energy can be tuned by an external flux $\Phi_{\textrm{ext}}$ threading the SQUID and the system is described by the Hamiltonian in \eqref{eq:chain} with $M=1$.
From numerical diagonalization, we determine that four data qubits accurately represent the low-lying states of one multilevel transmon.
The four lowest eigenenergies obtained from numerically exact diagonalization are shown in \figref{fig:single_transmon_VQD}(c).

We then employ the variational quantum eigensolver (VQE) and the variational quantum deflation (VQD) algorithms to estimate the same four lowest eigenenergies (see Sec.~\ref{sec:methods} and Appendix \ref{app:sec:vqe-circuits}).
Two layers of the four-qubit instance of the ansatz shown in \figref{fig:single_transmon_VQD}(b) are used as the parametrized unitary $\hat{U}_{\boldsymbol{\theta}}$.
The ansatz is inspired by the circuit template used in Ref.~\cite{Sousa2006} and by the circuit structure realizing ``quantum circuit Born machines'' \cite{Benedetti2019}.
It comprises an initial layer of single-qubit rotation gates followed by sequences of M\o lmer-S\o rensen ($XX$) gates and another layer of single-qubit rotation gates.
The use of $XX$ gates highlights the applicability to trapped ion processors, for which these gates are native two-qubit operations.
The circuit template is chosen based on the high ``expressibility'' of the ansatz, which has been shown to be helpful for modest-sized systems in which the general structure of eigenstates is not well-understood \cite{Sim2019}.
For parameter updates, we use the L-BFGS-B optimization routine \cite{Byrd1995}.
A layer-wise training strategy for optimizing the circuit parameters is used, sequentially optimizing sets of parameters corresponding to layers in a multi-layered parameterized quantum circuit.
The energy errors shown in \figref{fig:single_transmon_VQD}(c) are computed as the absolute difference between the variational values and those determined numerically via exact diagonalization.
The energies are estimated to high accuracy with errors falling below a threshold of $10$\,MHz, signalling that the VQE and VQD errors are small with respect to the characteristic energy scale of about 8\,GHz for the transmon.

\begin{figure}[t]
\centering
\includegraphics[scale=1.]{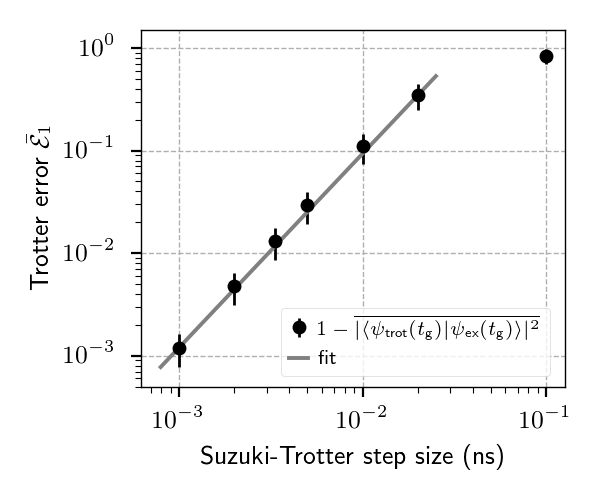}
\caption{
Digital quantum simulation of a transmon bit-flip gate operation.
The Trotter error $\bar{\mathcal{E}}_1$ is shown versus single step Trotter time.
For each Trotter step size, we perform Suzuki-Trotter evolution under the DRAG pulse for time $t_{\textrm{g1}}=85.32$\,ns for $300$ Haar random initial states spanned by $\{{\ket{0},\ket{1}}\}$. 
We obtain an average gate fidelity of $\bar{F}_1=99.68\pm 0.15\%$ and an average Trotter error of $\bar{\mathcal{E}}_1=0.12\pm 0.04\%$ at $1000$ Trotter steps. 
The polynomial fit shows a scaling behaviour of $\bar{\mathcal{E}}_1 \approx A_1 \times \Delta t^{\nu_1}$, with $\nu_1=1.89$ and $A_1=566$. 
}
\label{fig:NOTgate_exact}
\end{figure}

In addition to aiding in the design and optimization of the flux dispersion, the variational results can be used to estimate depolarization and decoherence times.
Here we show how dielectric loss, which is one of the dominant depolarization channels for transmons \cite{wang2015surface, krantz2019quantum}, can be estimated.
The relaxation time $T_1^\text{diel}$ from the first excited to the ground state is given by Fermi's Golden Rule:
\begin{equation}
    \label{eq:T1}
    \frac{1}{T_1^\text{diel}} = \frac{S_\text{diel}\left(\omega_{1}\right)}{\hbar^2} \left| \bra{e} 2e\hat{N} \ket{g} \right|^2,
\end{equation}
where $\ket{g}$ and $\ket{e}$ are the ground and excited state wave functions of the transmon circuit, respectively, $S_\text{diel}\left(\omega\right)$ is the noise spectral density of the lossy capacitance, and $\omega_{1}$ is the transition frequency between the ground and excited state \cite{smith2020superconducting,scqubits}.
We follow Ref.~\cite{scqubits} and use the ohmic-like noise spectral density
\begin{equation}
    S_\text{diel}\left(\omega\right) = \frac{2\hbar}{\left(C_\text{sh}+C_\text{J}\right)\times10^6} \left( \frac{\omega}{2\pi\times 6\text{GHz}} \right)^{0.7},
\end{equation}
assuming that the effect of blackbody radiation is negligible at the transition frequency \cite{nguyen2019high,smith2020superconducting}.
We now use the variationally determined ground and excited state to evaluate \eqref{eq:T1} and find $T_{1,\text{est}}^\text{diel} = 16.150\,\mu\text{s}$.
If we were to use the numerically exact eigenstates, we would find a lifetime that is longer by $1.2\,$ns, which corresponds to a relative error of $7.3
\times 10^{-5}$. 
Such an error is negligible for all practical purposes.
On a quantum computer, the expectation value in \eqref{eq:T1} could be determined with a SWAP test \cite{buhrman2001quantum}:
Using the appropriate variational circuit, one register of four data qubits is prepared in $\ket{e}$.
On a second register, the variational circuit for the state $\ket{g}$ is applied, followed by the Pauli-decomposed number operator $\hat{N}$ (provided in Appendix \ref{app:sec:cooper_pair_basis}).
The expectation value between the states is then computed with a SWAP test, which requires a single ancilla data qubit.
Other depolarization and dephasing processes of the circuit -- and dipole matrix elements in general -- can be readily determined on a quantum computer using similar procedures.

The variational eigenstates are also needed for simulating quantum gates for the hardware transmons.
As an example, we simulate the bit-flip gate, which transforms the computational state $\ket{0}$ to $\ket{1}$ and vice versa. 
Since the transmon is a multi-level system with small anharmonicity between its successive energy levels (see \figref{fig:single_transmon_VQD}(c)), resonantly driving the two lowest energy levels can populate multiple higher levels, resulting in undesired leakage. 
To address this challenge, the derivative removal by adiabatic gate (DRAG) scheme \cite{motzoi2009simple,gambetta2011analytic,de2015fast} is used (Appendix \ref{app:sec:drag}).
Suzuki-Trotter time evolution of the DRAG gate with $1000$ Trotter steps is numerically simulated and the population in the ground state is reliably transferred to the excited state within a gate time of $t_\text{g1} = 85.32\,\text{ns}$, minimizing leakage to undesired states.
The numerically exact eigenstates are used for all Suzuki-Trotter simulations.

To evaluate the performance of the Trotterized bit-flip gate, we prepare $300$ Haar random initial states, sampled from the Bloch sphere spanned by $\{\ket{0},\ket{1}\}$, and evolve them both numerically exactly and under the Trotterized digital quantum circuit for various Trotter step sizes.
The average gate fidelity, as defined in Appendix~\ref{app:sec:gates_fidelity}, is found to be $\bar{F}_1= 99.68\pm0.15\%$.
It is limited by remaining population in the ground state and can be increased by further optimization of the DRAG parameters.
The average Trotter error at the gate time $t_\text{g1}$ is $\bar{\mathcal{E}}_1=0.12\pm 0.04\%$ at $1000$ Trotter steps.
The Trotter error decreases polynomially with step size $\Delta t$ (see \figref{fig:NOTgate_exact}).
The extracted error scaling $\bar{\mathcal{E}}_1 \propto \Delta t^{\nu_1}$ with $\nu_1=1.89$ is superlinear and thus more favorable than the linear theoretical upper bound given by the Suzuki-Trotter approximation, \eqref{eq:Trotterized_unitary}.
Notably, potential additional errors caused by the time dependence of the driven Hamiltonian are not sufficient to push the error scaling beyond the time-independent theoretical bound.

\subsection{Two-transmon cell design}
\label{subsec:two-transmon}

\begin{figure}[t]
\centering
\includegraphics[scale=0.4]{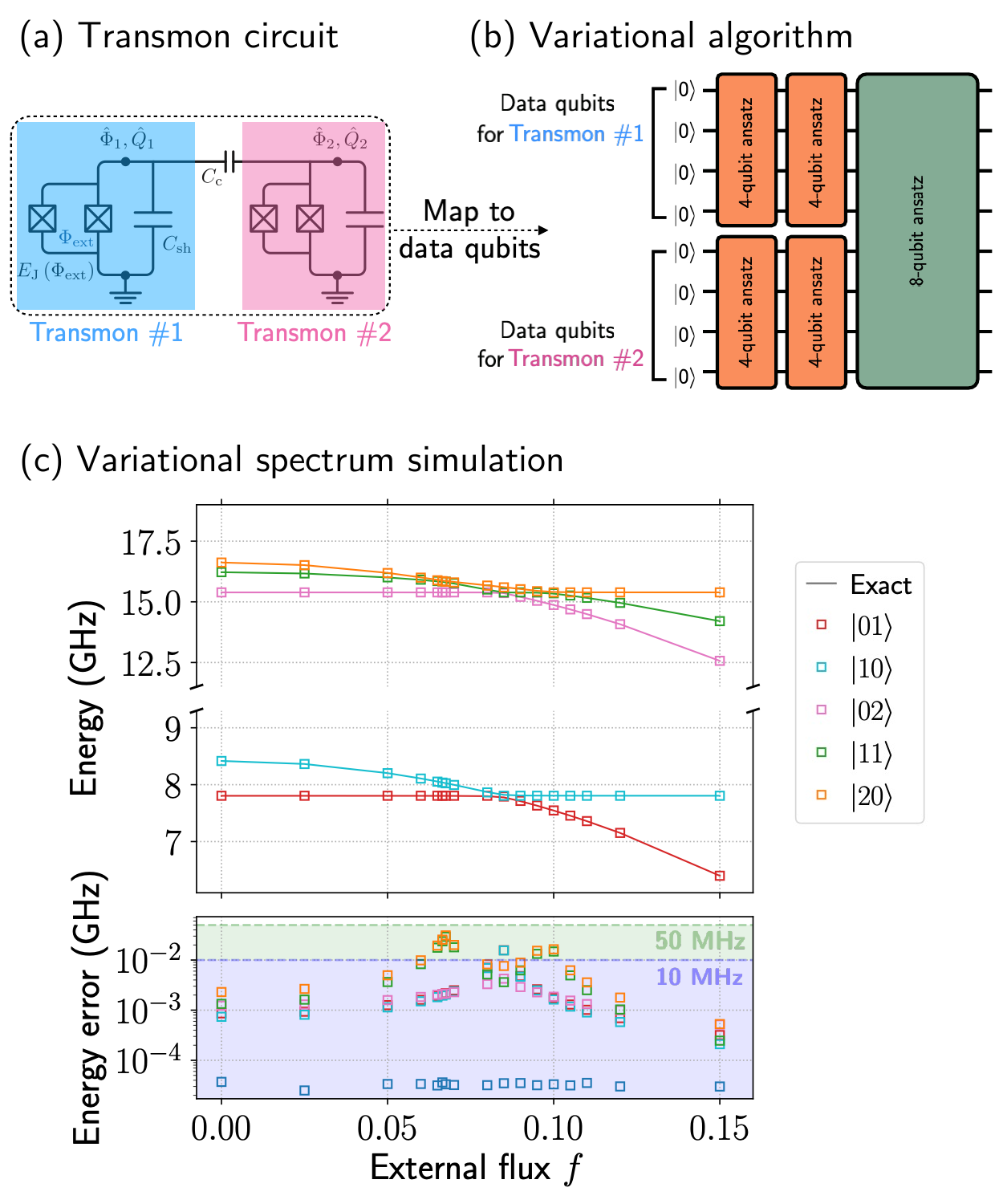}
\caption{
(a) Circuit diagram of two frequency tunable transmons coupled via a capacitor.
(b) The circuit is mapped to eight data qubits, with four data qubits each approximately representing one of the interacting transmons.
The variational ansatz leverages the four-qubit ansatz from the single-transmon simulation to seed the optimization over the full system.
(c) The lowest five excited state energies, with respect to the ground state energy, are shown as a function of the external flux $f_1=\Phi_{\text{ext,1}}/\Phi_0$ threading the left transmon, while the second external flux $f_2=0$ is held constant.
The solid lines are the exact diagonalization result, where the basis of the circuit variables $\hat{\Phi}_i$ and $\hat{Q}_i$ is truncated to $d=16$ states for both circuit nodes $i=1,2$.
The avoided level crossing between the states $\ket{11}$ and $\ket{20}$ at $f_1\approx 0.6$ corresponds to the operating point of the CPHASE gate.
Colored markers indicate the VQE- and VQD-optimized excitation energies.
The absolute energy differences between the VQE or VQD calculations and the exact diagonalization are shown in the plot below. 
The blue and green shaded regions indicate regions of error below $10$ MHz and $50$ MHz, respectively.
}
\label{fig:two_transmon_VQD}
\end{figure}

The quantum processor design also requires two-qubit simulations en route to a larger processor size.
Although we have found the energy spectrum of an individual transmon, it has to be updated when two or more transmons are coupled together as shown in \figref{fig:two_transmon_VQD}(a).
Here we set Josephson energies of $E_\text{J}/h =22$\,GHz (left transmon) and $E_\text{J}/h =19$\,GHz (right transmon), capacitances of $C_\text{sh}+ C_{\textrm{J}}=91$\,fF for both transmons, and a coupling capacitance $C_\text{c} =0.5$ fF.
The Hilbert space dimension has now increased and eight data qubits are required for quantum simulation.
Since the two transmons interact via the coupling capacitor, avoided level crossings are expected to appear in the energy spectrum.
These are essential for the design of two-transmon quantum gates.

We estimate the eigenenergies of the two-transmon system by again applying the VQE and VQD algorithms.
The variational ansatz comprises two blocks of the four-qubit ansatz that was used for the single transmon, followed by an eight-qubit ansatz that is constructed  accordingly (see \figref{fig:two_transmon_VQD}(b)).
Parameters for the two four-qubit blocks are first optimized assuming a system of individual \textit{uncoupled} transmons. 
These parameter settings are then used to initialize the corresponding parameters in the full ansatz for the coupled transmons. 
The VQD estimate of the first five excited states with respect to the VQE-optimized ground state energy are shown in \figref{fig:two_transmon_VQD}(c) along with the exact diagonalization result.
The normalized external flux $f_1$ (left transmon) is swept from $0$ to $0.15$ while $f_2 = 0$ (right transmon) is fixed.
In the regime where the two transmons are far detuned, the states are labelled as $\ket{01},\ket{10}, \ket{02},\ket{11},\ket{20}$ in order to keep track of how the respective states change and interact at the avoided level crossings.
Absolute energy differences between the VQD estimates and exact diagonalization are found to be below $50$\,MHz (green shaded region) around the avoided level crossings (see \figref{fig:two_transmon_VQD}(c)), as compared to $10$\,MHz for the single-transmon simulation.
We infer that variational optimization is harder when the state extends over a larger number of data qubits.

After performing the VQE and VQD optimizations, the initial states that are required for two-qubit gate simulations have been found.
As a proof-of-principle numerical experiment, we carry out digital simulation of the CPHASE gate, which features little leakage to undesired states.
There are two means to realize the gate: one via an adiabatic sweep of the external flux $f_1$ \cite{dicarlo2009demonstration} and another via an instantaneous (or diabatic) flux sweep \cite{dicarlo2010preparation}.
Here we take the latter approach and determine the gate parameters as follows:
We prepare the $\ket{11}$ state, suddenly tune $f_1$ close to the avoided level crossing, let $\ket{11}$ and $\ket{20}$ interact for some time, and suddenly change $f_1$ back to the initial point, imparting a phase on $\ket{11}$.
Post-rotations on the states remove spurious dynamical phases, resulting in a CPHASE operation.
Both numerically exact simulation and Suzuki-Trotter evolution with $50,000$ Trotter steps are performed, and we find that the gate is completed after an evolution time of $t_\text{g2} = 104.64\,\text{ns}$.

\begin{figure}[tb]
\centering
\includegraphics[scale=1.0]{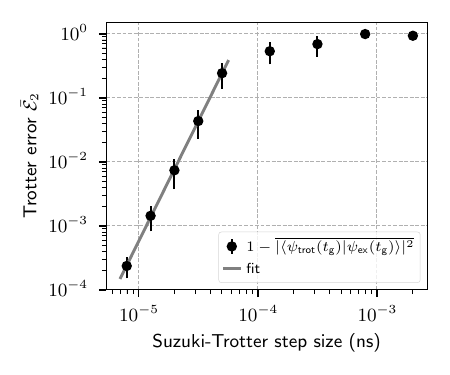}
\caption{
Digital quantum simulation of a two-transmon CPHASE gate operation.
The Trotter error $\bar{\mathcal{E}}$ is shown versus Trotter step size $\Delta t$.
We obtain an average gate fidelity of $\bar{F}_2=99.5\pm0.1\%$\ over $900$ Haar random initial states spanned by $\{{\ket{00},\ket{01},\ket{10},\ket{11}}\}$.
For each step size, we perform the CPHASE Suzuki-Trotter evolution over a gate time of $t_{\textrm{g2}}=104.64$\,ns for $100$ Haar random initial states spanned by the same computational basis.
At $50,000$ Trotter steps, the average Trotter error is $\bar{\mathcal{E}}_2=0.7\pm0.4\%$. 
The polynomial fit shows a scaling behaviour of $\bar{\mathcal{E}}_2 \approx A_2 \times \Delta t^{\nu_2}$, with $\nu_2=3.75$ and $A_2=3.2\times10^{15}$. 
}
\label{fig:two-transmon_CPHASE_gate}
\end{figure}

The performance of the numerically exact gate evolution as compared to the ideal CPHASE operation is quantified by the average gate fidelity over 900 Haar random initial state samples.
It is given by $\bar{F}_2 = 99.5\pm 0.1\%$ and is comparable to state of the art experimental demonstrations \cite{google_quantum_supremacy}.
The average Trotter error over 100 Haar random states amounts to $\bar{\mathcal{E}}_2=0.7\pm0.4\%$ for $50,000$ Trotter steps.
The Trotter error decreases polynomially with the Trotter step size $\Delta t$.
The simulation data in \figref{fig:two-transmon_CPHASE_gate} shows an error scaling $\bar{\mathcal{E}}_2 \propto \Delta t^{\nu_2}$ with $\nu_2=3.75$ for sufficiently small step sizes.
The polynomial scaling onset of the Trotter error $\bar{\mathcal{E}}_2$ appears for step sizes below $10^{-4}\,\text{ns}$, six orders of magnitude below the gate time.
For the single-qubit gate simulation, the step size only needed to be less than four orders below the gate time.
A likely cause for the discrepancy lies in the diabatic turn-on of the CPHASE gate interaction, whereas the single qubit pulse is smooth and slow in the rotating frame.
Besides, the Trotter error scales with almost fourth order in the step size.
This scaling factor is much larger than both the linear upper bound given by the Suzuki-Trotter approximation and the almost quadratic scaling of the single-transmon gate Trotter error. 
Therefore, the Trotter error can be reduced quickly by decreasing the step size once the polynomial onset has been reached.

\subsection{Scaling to a multi-transmon sub-module}
\label{subsec:multi-qubit}

It remains to show that our hardware simulation approach is scalable.
In the case of the transmon chain sub-module in \figref{fig:intro}(b-c), variational algorithms would shed light on unwanted cross-talk arising from spurious capacitances between any or all transmons in the sub-module.
Such interactions can be gleaned from the static energy spectra computed using variational quantum algorithms.
Here we study the performance of the VQE and VQD algorithms on a linear chain of up to four nearest-neighbor capacitively coupled transmons.
With the encoding of four data qubits per transmon, we thus simulate variational algorithms on up to 16 data qubits.
This is the largest size we can simulate with our computing resources in a reasonable runtime, and it is on par with other large numerical VQE simulations performed to date \cite{lee2018generalized, grimsley2019adaptive, dallaire2020application}.

We choose a variational ansatz that is different from the one used for the 1- and 2-transmon simulations.
Specifically, we employ the Josephson sampler ansatz from Ref.~\cite{geller2018sampling}, which was 
chosen to reduce the number of optimizer iterations and the evaluation time per circuit instance as the transmon number is increased.
It is constructed from CNOTs between all nearest neighbor pairs of qubits, which requires two layers of parallel CNOT gates.
Single-qubit $Y$ and $Z$ gates are performed before each entangling layer.
The circuit is repeated for multiple layers, and it is tiled to successively entangle all blocks of four data qubits that each encode a transmon (see Appendix~\ref{app:sec:vqe-circuits}).
The Josephson sampler ansatz (JSA) has been found to exhibit high expressibility and entangling capability \cite{Sim2019} while featuring reduced resource requirements as compared to our previous ansatz.
It requires fewer two-qubit gates and fewer variational parameters, as the two-qubit gates are not parameterized.
We note that the ansatz is particularly suitable for quantum computing platforms with high-fidelity nearest neighbor gates, such as superconducting processors.

Using the Josephson sampler ansatz, we perform VQE and VQD simulations of up to five lowest-lying energy eigenstates for chains of up to four transmons.
The transmon circuit diagrams are the same as before, with junction energies of $E_\text{J}/h =22$\,GHz for the first transmon and $E_\text{J}/h =19$\,GHz for all successive transmons.
Therefore, transmons 2, 3, and 4 are strongly interacting at their fixed flux of $0\,\Phi_0$, forming eigenstates that are spread over all of the data qubits that encode them.
The first excited state of transmon 1 has a larger energy and it is tuned into resonance with the others as its variable flux is increased.
By choosing a system with highly interacting modes, we have created a particularly difficult benchmark problem for the variational algorithms.

\begin{figure}[htb]
\centering
\includegraphics[scale=0.5]{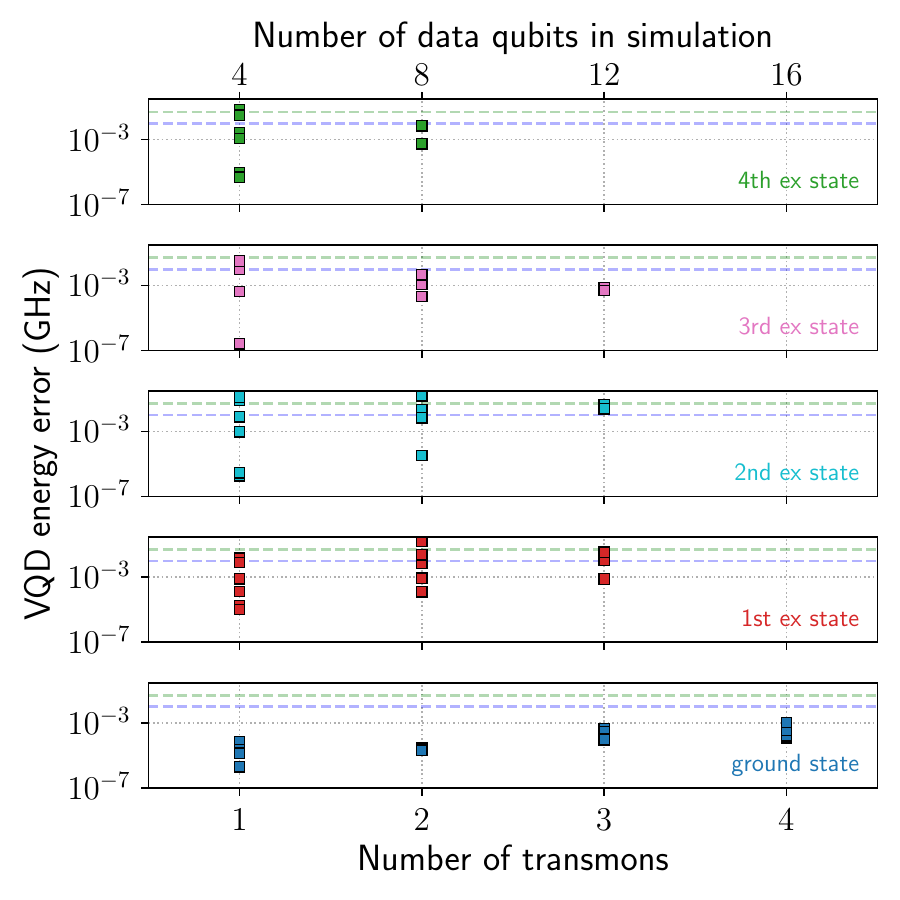}
\caption{
Scaling of the variational simulations to larger system sizes.
The absolute energy difference between the VQE or VQD calculations and the exact diagonalization is shown for the five lowest lying states of an increasingly larger system.
The system size ranges from a single transmon to a linear chain of four transmons, which takes 16 data qubits to simulate.
For the largest system size, only the ground state is simulated.
The vertically distributed points of the same color correspond to simulations of the same eigenstate at several flux points between $0.0\,\Phi_0$ and $0.2\,\Phi_0$.
The green (blue) horizontal line corresponds to the previously identified error threshold of 50\,MHz (10\,MHz).
The typical error for a VQD excited state simulation does not exhibit an increasing trend with the system size, and all VQE ground state energies stay below the error thresholds.
}
\label{fig:scalability}
\end{figure}

The error of the variational VQE and VQD simulations with respect to the exact diagonalization is shown in \figref{fig:scalability}.
The ground state and up to four excited states are simulated.
For the largest system size composed of four transmons, we limit the computations to the ground state, as the excited states exceed the capabilities of the employed classical computing resources.
As we are primarily interested in how well the method scales, we do not show the full flux dispersion but rather cluster the errors for a few different flux points.
Remarkably, the excited state data do not show a significant increase in the variational error as the system size increases.
Moreover, most of the errors are below the previously identified threshold of 50\,MHz.
For the ground state, we observe a slight increase in the error but it stays well below the threshold for all system sizes.
Thus, our numerical simulations suggest that our method could be scalable to transmon processor design beyond the capabilities of classical computers.

Relevant metrics for determining the experimental suitability of a variational algorithm are the number of data qubits, circuit depth, and gate count required to simulate an $M$-transmon processor sub-module.
That is because for NISQ computers, errors accumulate with the number of gates and the circuit depth.
Based on our numerical VQE and VQD simulations, we propose a circuit structure that performed well for two transmons and may be naturally extended to an $M$-transmon chain.
Each transmon is again encoded in four data qubits such that the total data qubit count scales linearly with $M$.
We apply one layer of the four-qubit template of the Josephson sampler ansatz for each transmon. To entangle the subsystems, we apply two layers of an eight-qubit template. This circuit structure performed well for modeling two capacitively-coupled transmons, so we consider this as a fundamental building block and refer to it as a JSA block.
It is then used to extend to the case of $M>2$ transmon chain.
Appendix~\figref{fig:ansatz_circuit} details the JSA construction as well as the 3- and 4-transmon ansatze.

There are three potential ways to scale up the JSA. 
These are the triangle-layer ansatz, brick-layer ansatz and reversed multi-scale entanglement renormalization ansatz (MERA) \cite{vidal2008class}, respectively.
We have outlined the exact numbers of CNOTs and circuit layers in Appendix \ref{app:sec:vqe-circuits} and the circuits are shown in Appendix~\figref{fig:ansatz_circuit}.
The triangle-layer ansatz can be seen as a big triangular configuration composed of JSAs, covering the entire circuit depth from left to right. 
The base of the triangle is where the quantum circuit begins and its tip is at the rightmost end of the circuit.
The brick-layer ansatz circuit consists of JSAs located alternatively across circuit layers. 
Suppose a JSA is located across qubit 1 and 2 in the first layer.
That means, in the second layer, another JSA is spanned across qubit 2 and 3.
In both cases, we assume that the circuit depth is the same.
Lastly, the reversed MERA is an inverted MERA circuit, which can be seen as a bifurcating tree structure with a JSA node splitting into two sub-JSA blocks. 
For all the three cases, we find that the gate count and depth of the variational circuit scale favorably with the number of transmons $M$.
For both the triangle-layering and the brick-layering ansatz, the circuit depth scales with the number of $M$ transmons in the linear chain $\mathcal{O}(M)$, while the number of 2-qubit CNOT gates scales as $\mathcal{O}(M^2).$
For the reversed MERA, the circuit depth scales as $\mathcal{O}(\log_2 (M/2))$ and the number of CNOTs as $\mathcal{O}(M).$

The dynamical gate simulations developed in this work are tailored to intermediate-term or long-term error-corrected quantum computers, and they are beyond the reach of near-term devices. 
The reason is the large number of $304$ and $152$ entangling gates (upper bound) per single Trotter step to simulate the single-qubit bit-flip gate and the two-qubit CPHASE gate, respectively.
The number of gates in the single-qubit simulation is larger than in the two-qubit gate simulation because of the DRAG scheme, where the Hamiltonian is a function of projection operators instead of sparse matrices (Appendix \ref{app:sec:drag}).
The gate upper bounds are estimated by cancelling adjacent gates in the Suzuki-Trotter evolution, exploiting commutation relations, and simplifying commonly occurring patterns as described in \cite{nicolas2019}.
Afterwards, the remaining CNOT gates are counted.
A total of $1000$ and $50,000$ Trotter steps were applied to digitally quantum simulate the single-transmon and the two-transmon gate, attaining favourable Trotter errors $\bar{\mathcal{E}}_1=0.12\pm0.04\%$ and $\bar{\mathcal{E}}_2=0.7\pm0.4\%$, respectively.
Therefore, a total of $3.04\times 10^5$ CNOTs is required for the bit-flip simulation and $7.6\times 10^6$ CNOTs for the CPHASE simulation.
The gate count can be reduced by tolerating a higher Trotter error or by applying further gate count reduction techniques \cite{childs2019faster,campbell2019random}.
Nevertheless, such simulations likely remain beyond the capabilities of near-term processors.

To perform transmon gate simulations using NISQ devices, it would be necessary to move away from pure Suzuki-Trotter evolution.
In fact, multiple approaches have been developed to perform time evolution of quantum systems with near-term variational quantum algorithms.
The variational quantum simulation (VQS) algorithm in Ref.~\cite{li2017efficient} replaces each step of the time evolution by a variational optimization, thus faithfully reproducing the time evolution and even outperforming the Suzuki-Trotter method.
Although time-dependent Hamiltonians such as the bit-flip gate discussed in our work are not covered, it appears straightforward to update the Hamiltonian in each time step.
As the time dependence of the bit-flip pulse is smooth and did not lead to a violation of the time-independent error bound in our simulations, we expect that the pulse discretization error would also be small for the VQS algorithm.
The large number of variational optimizations that need to be performed, however, is a drawback of VQS.
A faster runtime can be achieved by variational fast forwarding algorithms, which learn from small Trotter steps or static Hamiltonians to predict the evolution for longer times \cite{cirstoiu2020variational, commeau2020variational}.
While fast forwarding techniques are particularly useful for the time-independent Hamiltonian in the CPHASE gate simulation, their advantage is less clear for the time-dependent drive in the bit-flip gate.
While the focus of the present work lies on the performance of the Suzuki-Trotter method, which holds promise for long-term quantum computing, exploring the use of near-term variational time evolution techniques presents a promising line of future work.

\section{Conclusion}
\label{sec:conclusion}
The best quantum processors today are already too large to be simulated on any existing classical machine. Their growth has been approximately exponential since the first transmon device in 2008 \cite{schreier2008suppressing}.
If one hopes to capture the behavior of design primitives and sub-modules that grow with processor size, future quantum processors will have to be simulated on quantum computers.
In this work, we have developed and simulated a detailed set of hybrid classical-quantum algorithms that can be used for quantum computer-aided design and assessed their scalability. 
With the use of the variational quantum eigensolver and variational quantum deflation algorithm for near-term quantum processors, we numerically simulated quantum circuits with up to 16 data qubits that encode up to four capacitively coupled transmons.
Errors incurred from running these algorithms are small with respect to the dominant system energy scale.
While the results presented here assume a noiseless quantum processor, protocols have been developed to mitigate the effect of noise in practice: For example, zero-noise extrapolation \cite{li2017efficient,temme2017error} and probabilistic error cancellation \cite{temme2017error,endo2018practical} do not incur any additional qubit cost and only require relatively simple classical post-processing.
For a review of error mitigation techniques, we refer to Refs. \cite{bharti2021noisy,endo2021hybrid}.
Based on the variational simulations, we were able to prepare transmon eigenstates and pinpoint avoided energy level crossings which are then used to simulate dynamical one- and two-transmon hardware gates with state-of-the-art fidelities.
While Trotter errors scale favorably with the Suzuki-Trotter step size, the gate counts suggest that these dynamical simulations need to be performed on long-term quantum hardware.
As an alternative approach that is more amenable to near-term hardware, we laid out a path to perform dynamical simulations with variational quantum simulators.
Our work indicates that quantum computer-aided design enables exploration of circuit parameters within the quantum processor sub-module without fabricating and re-designing physical hardware, thereby saving time and cost. 
In order to have further design control, one may combine our proposed scheme with classical computation methods for superconducting circuit design and simulation \cite{menke2019automated,li2020towards,di2019efficient}, thereby benefiting from both the classical and quantum realm in a single design pipeline.
In addition, our methodology of encoding multi-level quantum hardware in data qubits and investigating static and dynamic properties with near- and long-term quantum algorithms provides a general framework that is readily applicable to trapped ion, neutral atom, photonic, and other systems.

\section*{Acknowledgements}
We acknowledge J. Braum\"uller, M. Degroote, I.D. Kivlichan, M. Krenn, M. Kjaergaard, A. Di Paolo, and G.O. Samach for valuable discussions.
We also thank two anonymous reviewers for their helpful comments on the manuscript.
T.H.K., T.M., and A.A.-G. acknowledge funding from Intel Research and from Dr. Anders G. Fr{\o}seth.
T.M. and A.A.-G. also acknowledge the Vannevar Bush Faculty Fellowship under contract ONR N00014-16-1-2008.
S.S. is supported by the Department of Energy Computational Science Graduate Fellowship under grant number DE-FG02-97ER25308.
A.A.-G. also acknowledges support from the Canada 150 Research Chairs Program, the Canada Industrial Research Chair Program, and from Google, Inc. in the form of a Google Focused Award.
This work was also supported by the U.S. Department of Energy under Award No. DE-SC0019374.
The Niagara supercomputer at the SciNet HPC Consortium \cite{loken2010scinet, ponce2019deploying} was used for a part of the computations in this work.
SciNet is funded by: the Canada Foundation for Innovation; the Government of Ontario; Ontario Research Fund - Research Excellence; and the University of Toronto.

\appendix

\section{Physical qubit Hamiltonians in the charge number bases}
\label{app:sec:cooper_pair_basis}
The one-transmon and two-transmon Hamiltonians, which are the $M=1$ and $M=2$ special cases of \eqref{eq:chain}, are given by
\begin{align}
    \hat{H}_\text{1-transmon} = & \, 4 E_\text{C} \hat{N}^2 - 2E_\text{J} |\cos (2\pi f)|\cos \hat{\varphi},
    \label{eq:one-qubit} \\
    \hat{H}_\text{2-transmon} = & \, 4E_\text{C} (1+\xi)^{-1} (\hat{N}_1 ^2 + \hat{N}_2 ^2 + 2 \xi \hat{N}_1 \hat{N}_2)
      \label{eq:two-qubit} \\
      & - 2E_{\text{J},1} |\cos (2\pi f_1)|\cos \hat{\varphi}_1 \nonumber \\
      & - 2E_{\text{J},2} |\cos (2\pi f_2)|\cos \hat{\varphi}_2. \nonumber
\end{align}
The parameter $E_\text{C} = e^2/2(C_{\textrm{sh}}+C_\text{J})$ is the charging energy and  $\xi = C_\text{c} /(C_\text{c} + C_{\textrm{sh}} + C_{\textrm{J}})$.
We note that the labelling $1,2$ in the Hamiltonian $\hat{H}_\text{2-transmon}$ is not limited to nearest-neighbour transmons, but the two transmons can in principle be separated by some other transmons within the same sub-module \footnote{In this setup, the form of the Hamiltonian still holds, but the prefactor $\xi$ changes accordingly.}.

In order to represent the above Hamiltonians in discrete charge number bases, let us look at the quantum mechanical operators corresponding to an active node in the transmon circuit seen in \figref{fig:single_transmon_VQD} and \figref{fig:two_transmon_VQD} satisfy the commutation relation $[\hat{\varphi}_j, \hat{N}_k]= i \delta_{jk}$.
One can show the resulting relations
\begin{equation}
\label{Eq:raising_lowering}
	[e^{i\hat{\varphi}_j},\hat{N}_j]=-e^{i\hat{\varphi}_j}, \,\,\, e^{\pm i\hat{\varphi}_j}\ket{n_j}=\ket{n_j \pm 1},
\end{equation}
where $\ket{n_j}$ are the eigenstates of $\hat{N}_j$.
We notice that the operators $e^{\pm i\hat{\varphi}_j}$ are similar to the usual bosonic creation and annihilation operators, without the square root pre-factor. They are, in fact, the Susskind-Glogower phase operators \cite{susskind1964quantum}.
Hence, the circuit Hamiltonians can be rewritten with
\begin{equation}\label{Eq:n_operator}
	\hat{N} = \sum_{n=0}^{d-1} \left(n-\frac{d}{2} \right) \ketbra{n}{n},
\end{equation}
\begin{equation}\label{Eq:cosine_operator}
	\cos \hat{\varphi} = \frac{1}{2} \sum_{n=0}^{d-2} (\ketbra{n}{n+1}+\ketbra{n+1}{n}).
\end{equation}

In general, the number of Cooper pairs can take on infinitely many integer values.
However, for practical purposes, we are only interested in the low lying energy states.
Therefore, we can truncate the Hilbert space as described in the main text by limiting the number of basis states to $d= 2^k, \, k \in \mathbb{N}$.
The number of data qubits that are used in the quantum simulation of a transmon is then given by $k$.

\section{Encoding of quantum operators into Pauli strings}
\label{app:sec:encoding}
\begin{figure}[t]
\centering
\includegraphics[scale=0.32]{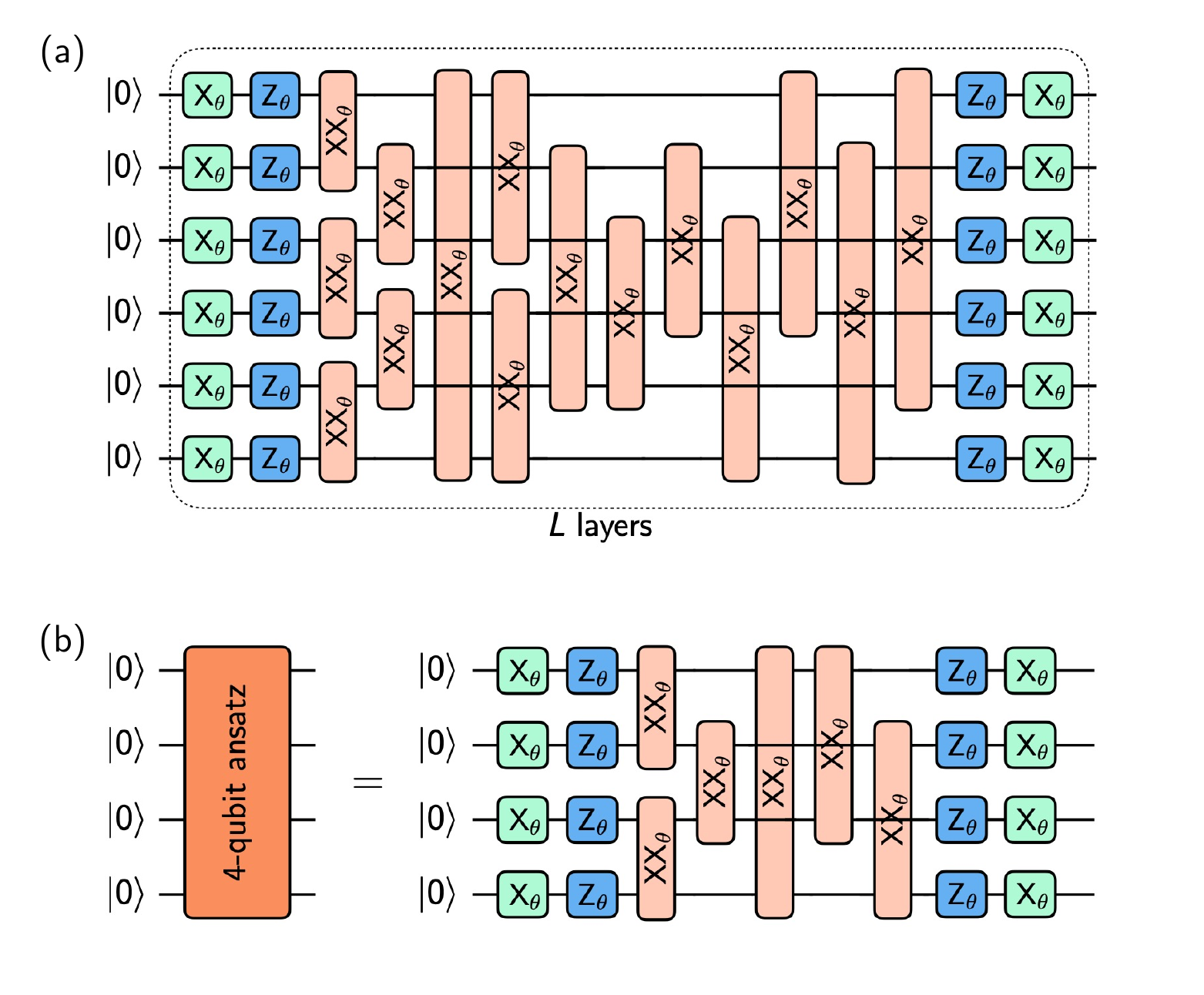}
\caption{Parameterized quantum circuits used to compute spectra of one and two transmons. (a) Six-qubit version of the parameterized quantum circuit structure used for VQE and VQD simulations of qubit designs investigated in the study of flux qubit designs. The dotted box around the circuit denotes a unit layer, which can be repeated $L$ times to potentially generate a more expressible circuit.
(b) Four-qubit version of the circuit structure used to simulate a single transmon. Two unit layers of this four-qubit ansatz were employed in simulations. 
}
\label{fig:vqe_circuit}
\end{figure}
\begin{table*}[htb]
\begin{center}
\begin{tabular}{|c|c|c|c|}
\hline
 $d=16$ & Std. Binary & Gray & Unary \\
\hline
 $\hat N$ 
& \begin{tabular}[c]{@{}c@{}}
$-0.5$ $I$ \\ 
$- 4.0$ $Z_{3}$\\ 
$- 2.0$ $Z_{2}$\\ 
$- 1.0$ $Z_{1}$\\ 
$- 0.5$ $Z_{0}$\\ 
\end{tabular}  
& \begin{tabular}[c]{@{}c@{}}
$-0.5$ $I$ \\ 
$- 4.0$ $Z_{3}$\\ 
$- 2.0$ $Z_{2}Z_{3}$\\ 
$- 1.0$ $Z_{1}Z_{2}Z_{3}$\\ 
$- 0.5$ $Z_{0}Z_{1}Z_{2}Z_{3}$\\ 
\end{tabular}
& 
\begin{tabular}[c]{@{}c@{}}
\\
$-4.0$ $I$  
$+ 4.0$ $Z_{0}$\\ 
$+ 3.5$ $Z_{1}$  
$+ 3.0$ $Z_{2}$\\ 
$+ 2.5$ $Z_{3}$  
$+ 2.0$ $Z_{4}$\\ 
$+ 1.5$ $Z_{5}$  
$+ 1.0$ $Z_{6}$\\ 
$+ 0.5$ $Z_{7}$  
$- 0.5$ $Z_{9}$\\ 
$- 1.0$ $Z_{10}$  
$- 1.5$ $Z_{11}$\\ 
$- 2.0$ $Z_{12}$  
$- 2.5$ $Z_{13}$\\ 
$- 3.0$ $Z_{14}$ 
$- 3.5$ $Z_{15}$\\ \\
\end{tabular}
\\

\hline

$\cos\hat\varphi$ 
&  \begin{tabular}[c]{@{}c@{}}
$+ 0.5$ $X_{0}$\\ 
$+ 0.25$ $X_{0}X_{1}$\\ 
$+ 0.25$ $Y_{0}Y_{1}$\\ 
$+ 0.125$ $X_{0}X_{1}X_{2}$\\ 
$+ 0.125$ $X_{0}Y_{1}Y_{2}$\\ 
$+ 0.125$ $Y_{0}X_{1}Y_{2}$\\ 
$- 0.125$ $Y_{0}Y_{1}X_{2}$\\ 
$+ 0.0625$ $X_{0}X_{1}X_{2}X_{3}$\\ 
$+ 0.0625$ $X_{0}X_{1}Y_{2}Y_{3}$\\ 
$+ 0.0625$ $X_{0}Y_{1}X_{2}Y_{3}$\\ 
$- 0.0625$ $X_{0}Y_{1}Y_{2}X_{3}$\\ 
$+ 0.0625$ $Y_{0}X_{1}X_{2}Y_{3}$\\ 
$- 0.0625$ $Y_{0}X_{1}Y_{2}X_{3}$\\ 
$- 0.0625$ $Y_{0}Y_{1}X_{2}X_{3}$\\ 
$- 0.0625$ $Y_{0}Y_{1}Y_{2}Y_{3}$\\ 
\end{tabular}
&  \begin{tabular}[c]{@{}c@{}}
$+ 0.5$ $X_{0}$\\ 
$+ 0.25$ $X_{1}$\\ 
$- 0.25$ $Z_{0}X_{1}$\\ 
$+ 0.125$ $X_{2}$\\ 
$- 0.125$ $Z_{1}X_{2}$\\ 
$+ 0.125$ $Z_{0}X_{2}$\\ 
$- 0.125$ $Z_{0}Z_{1}X_{2}$\\ 
$+ 0.0625$ $X_{3}$\\ 
$- 0.0625$ $Z_{2}X_{3}$\\ 
$+ 0.0625$ $Z_{1}X_{3}$\\ 
$- 0.0625$ $Z_{1}Z_{2}X_{3}$\\ 
$+ 0.0625$ $Z_{0}X_{3}$\\ 
$- 0.0625$ $Z_{0}Z_{2}X_{3}$\\ 
$+ 0.0625$ $Z_{0}Z_{1}X_{3}$\\ 
$- 0.0625$ $Z_{0}Z_{1}Z_{2}X_{3}$\\ 
\end{tabular}
& \begin{tabular}[c]{@{}c@{}}
\\
$+ 0.25$ $X_{0}X_{1}$ 
$+ 0.25$ $Y_{0}Y_{1}$\\
$+ 0.25$ $X_{1}X_{2}$ 
$+ 0.25$ $Y_{1}Y_{2}$\\ 
$+ 0.25$ $X_{2}X_{3}$ 
$+ 0.25$ $Y_{2}Y_{3}$\\ 
$+ 0.25$ $X_{3}X_{4}$ 
$+ 0.25$ $Y_{3}Y_{4}$\\ 
$+ 0.25$ $X_{4}X_{5}$ 
$+ 0.25$ $Y_{4}Y_{5}$\\ 
$+ 0.25$ $X_{5}X_{6}$ 
$+ 0.25$ $Y_{5}Y_{6}$\\ 
$+ 0.25$ $X_{6}X_{7}$ 
$+ 0.25$ $Y_{6}Y_{7}$\\ 
$+ 0.25$ $X_{7}X_{8}$ 
$+ 0.25$ $Y_{7}Y_{8}$\\ 
$+ 0.25$ $X_{8}X_{9}$ 
$+ 0.25$ $Y_{8}Y_{9}$\\ 
$+ 0.25$ $X_{9}X_{10}$ 
$+ 0.25$ $Y_{9}Y_{10}$\\ 
$+ 0.25$ $X_{10}X_{11}$ 
$+ 0.25$ $Y_{10}Y_{11}$\\ 
$+ 0.25$ $X_{11}X_{12}$ 
$+ 0.25$ $Y_{11}Y_{12}$\\ 
$+ 0.25$ $X_{12}X_{13}$ 
$+ 0.25$ $Y_{12}Y_{13}$\\ 
$+ 0.25$ $X_{13}X_{14}$ 
$+ 0.25$ $Y_{13}Y_{14}$\\ 
$+ 0.25$ $X_{14}X_{15}$ 
$+ 0.25$ $Y_{14}Y_{15}$\\ \\
\end{tabular}
\\
 
\hline

\end{tabular}
\end{center}
\caption{Qubit encodings (standard binary, Gray code, and Unary) of elementary operators used in this study, with a truncation of $d=16$. In our numerical experiments, we utilize the Gray code.}
\label{tab:enc16}
\end{table*}

In order to represent the superconducting circuit or any other type of $d$-level bosonic quantum hardware in terms of qubit states, we convert the eigenstates of the number operator $\hat{N}$ into the computational basis states of $k$ data qubits by representing the integer charge number in a preferred encoding \cite{veis2016quantum,mcardle2018quantum,nicolas2019}.
This implies a truncation of the physical space to the subspace spanned by $2^k$ Cooper pair numbers.
There are combinatorially many ways to map such a state space to a set of qubits. 
For all the numerical quantum simulation experiments presented throughout this work, we have employed the Gray code \cite{nicolas2019} due to its resource efficient representation of tridiagonal quantum matrix operators. 
After integer labeling, each level is encoded into a set of bits, which is then mapped to Pauli operators:
\begin{align}
	\ketbra{0}{1} &= (X+iY)/2 \, ,\nonumber\\
	\ketbra{1}{0} &= (X-iY)/2 \, ,\nonumber\\
	\ketbra{0}{0} &= (\mathbb{I}+Z)/2 \, ,\nonumber\\
	\ketbra{1}{1} &= (\mathbb{I}-Z)/2 \, .\nonumber
\end{align}
Here, $X, Y, Z$ are the usual Pauli matrices and $\mathbb{I}$ is the identity. 
Encodings for the operators seen in Eqs.~(\ref{Eq:n_operator},\,\ref{Eq:cosine_operator}) truncated at $d=16$ are given in Table \ref{tab:enc16}.
In terms of required two-data-qubit operations, the operator $\hat N$ is most efficient in the standard binary representation, but the operator $\cos \hat \varphi$ is more efficient in the Gray representation. 
Overall, the Gray code is superior, assuming one performs the Trotterization in only one encoding. 
However, if one converts between the standard binary and Gray encodings, using the former for $\hat N$ and the latter for $\cos \hat \varphi$ may be a more efficient approach during Trotterization \cite{nicolas2019}.


\section{Variational ansatz for VQE and VQD simulations}
\label{app:sec:vqe-circuits}

We use the variational quantum eigensolver approach to find the static energy spectra of superconducting circuits on near-term quantum hardware.
Here we provide the parameterized quantum circuit templates used for all VQE and VQD simulations reported throughout this work.
We note that the template that was used for the 1- and 2-transmon simulations in Secs.~\ref{subsec:single-transmon},~\ref{subsec:two-transmon} (see \figref{fig:vqe_circuit}) is different from the template for the scalability simulations in \secref{subsec:multi-qubit}.

\begin{figure*}[t]
\centering
\includegraphics[scale=0.65]{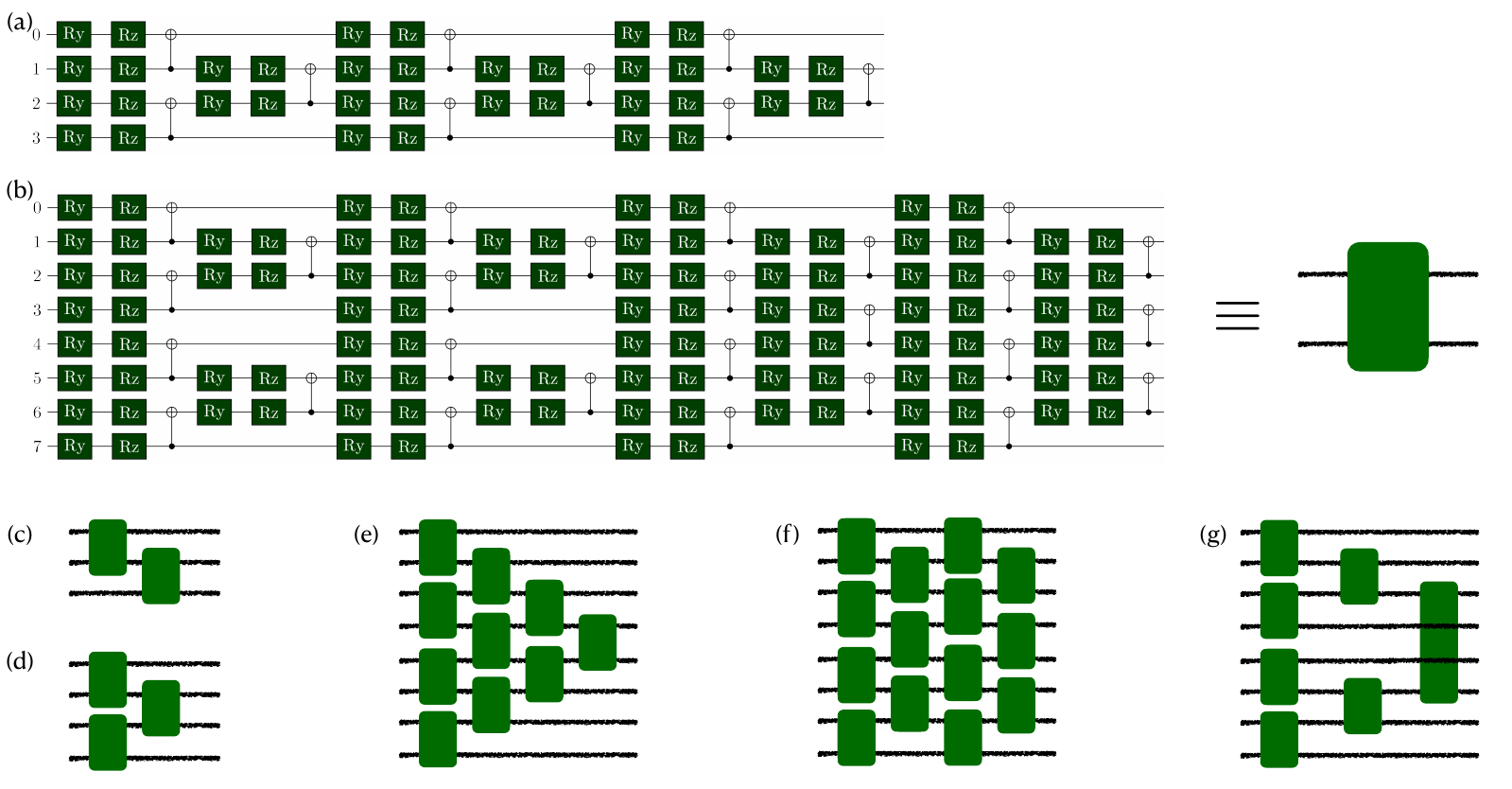}
\caption{Ansatz circuit for (a) single transmon (b) 2-transmon or single Josephson sampler ansatz block (JSA) (c) 3-transmon and (d) 4-transmon simulation, respectively.
Single black bold line in (c-g) represents four data qubit lines.
(e-g) show three potential scaling of the JSA shown in (b) for a large $M$-transmon simulation.
For both triangle-layering (e) and brick-layering ansatz (f), the circuit depth scales with the number of $M$ transmons in the linear chain $\mathcal{O}(M)$, while the number of 2-qubit CNOT scales as $\mathcal{O}(M^2).$
For the reversed MERA (g), the circuit depth goes with $\mathcal{O}(\log_2 (M/2))$, while the number of CNOT is $\mathcal{O}(M).$
In (g), we note that the central qubit 4 and 5 are not connected to the last JSA block.
}
\label{fig:ansatz_circuit}
\end{figure*}
The template for one and two transmons is shown in \figref{fig:vqe_circuit}(a).
The ``unit'' layer of this circuit, inspired by the circuit template used in Ref.~\cite{Sousa2006} and the circuit structure realizing ``quantum circuit Born machines'' (QCBM) \cite{Benedetti2019}, comprises an initial layer of single-qubit rotation gates followed by sequences of M\o lmer-S\o rensen ($XX$) gates and another layer of single-qubit rotation gates.
Each $XX$ entangling gate is defined as $XX_{ij}(\theta) = \text{exp}(-i \theta X_iX_j/2)$ for qubits $i$ and $j$.

We propose three alternative ways to scale up the ansatz for $M$ transmons using the JSA, see \figref{fig:ansatz_circuit}.
Since the resulting ansatze consist of many JSAs, we denote $\mathcal{J}$ to represent one JSA with sum of $52~\textrm{R}_y$ gates, $52~\textrm{R}_z$ gates, and $26~\textrm{CNOTs}$, with $24$ circuit layers.
Firstly, for triangle-layer ansatz (see \figref{fig:ansatz_circuit}(e)), we have the total JSA number, when $M$ is even, $n_{\mathcal{J},\Delta}= \sum_{k=1}^{M/2}k=(M^2 + 2M)/{8}$.
That means the number of CNOT is $n_{\textrm{CNOT},\Delta}=13(M^2+2M)/4.$
The subscript $\Delta$ denotes triangle-layer ansatz.
The circuit depth is $d_{\Delta}=12M.$
When $M$ is odd, we have $n_{\mathcal{J},\Delta}=(M^2+4M-5)/8$, corresponding to the total number of CNOT- $n_{\textrm{CNOT},\Delta}=13(M^2+4M-5)/4.$
The circuit depth then becomes $d_{\Delta}=(M+1)/2.$
Secondly, the brick-layer ansatz construction (see \figref{fig:ansatz_circuit}(f)) requires $n_{\mathcal{J},B}=\left(\lceil\frac{M}{4} \rceil \right).\frac{M}{2}+\left(\frac{M}{2}- \lceil\frac{M}{4} \rceil \right).\left(\frac{M}{2}-1 \right)$, for even $M$.
That means $n_{\textrm{CNOT},B}=26n_{\mathcal{J},B},$ and $d_B = 12M.$
The subscript $B$ denotes the brick layering.
For odd $M$, we have $n_{\textrm{CNOT},B}=13(M^2-1)/2,$ and $d_B =12(M+1).$
Lastly, for a reversed multi-scale entanglement renormalization ansatz (MERA) construction (see \figref{fig:ansatz_circuit}(g)), we can only have even $M$ and thus $n_{\textrm{CNOT},RM}=26(M-1)$ while $d_{RM}=24(2\log_2 ({M}/{2})-1).$
Here, $RM$ stands for reversed MERA.

\section{Many-level transmon and the DRAG scheme}
\label{app:sec:drag}

Unitary gate operation for a single transmon system is non-trivial and it ought to be discussed briefly here.
In this section, we derive the driven transmon Hamiltonian that is simulated for the bit-flip operation.
For concreteness, we focus on the single-transmon bit-flip gate or NOT gate, which can be attained by driving the resonator line capacitively coupled to a transmon.
The way to realize the NOT gate in general is to drive the two lowest energy levels with an external field that is resonant with their energy difference.
However, the transmon level structure has weak anharmonicity (see \figref{fig:single_transmon_VQD}(c)). 
Simply driving the two lowest energy levels would populate higher energy levels, thus reducing the gate fidelity.
In order to overcome this issue, the derivative removal by adiabatic gate (DRAG) scheme \cite{motzoi2009simple,gambetta2011analytic,de2015fast} is applied and the corresponding result can be seen in the main text.
The idea of the DRAG scheme is to use two slowly varying quadrature controls to instantaneously cancel leakages to undesired levels such as transitions to second and third excited states. 

The combined system and drive Hamiltonian can be modelled ($\hbar=1$) as
\begin{eqnarray}
    \hat{H}_{\textrm{gate}}&=&\hat{H}_{1\textrm{-transmon}}+\hat{H}_{\textrm{resonator}}+\hat{H}_{\textrm{interaction}}+\hat{H}_{\textrm{drive}}\nonumber\\
        &=&\sum_{j=1}^{d-1}(j\omega_1 +\Delta_j)\ketbra{j}{j} + \omega \hat{b}^\dagger \hat{b}\\
        &+& \sum_{j=1}^{d-1}\left[ g_{j,j-1} \ketbra{j}{j-1}\hat{b}+ g_{j-1,j}\ketbra{j-1}{j}\hat{b}^\dagger\right]\nonumber\\
        &+& \epsilon(t) (\hat{b}+\hat{b}^\dagger),\nonumber
\end{eqnarray}
where $\omega_1$ is the energy gap between the ground and first excited state of the transmon and $\Delta_j$ are the energy level anharmonicities, with $\Delta_1=0$, $\Delta_2 = \omega_2 - 2\omega_1$, and $\Delta_n = \omega_n - n\omega_1$, where $\omega_j$ is the eigenenergy of the eigenstate $\ket{j}$, i.e. $\hat{H}_{\textrm{1-transmon}} \ket{j}=\omega_j \ket{j}$.
The frequency of the single mode resonator is given by $\omega$. 
The operator $\hat{b}^\dagger(\hat{b})$ is the bosonic creation (annihilation) operator of the resonator, while $g_{j,k}$ is the light-matter coupling strength between the resonator mode and a specific transmon level transition $\ketbra{j}{k}$.
The time-dependent external drive amplitude of the resonator quadrature is given by $\epsilon(t)$.
In the above notation, we take the ground state energy to be zero for the projector $\ketbra{0}{0}$.
We also define the projectors $\hat{P}_j = \ketbra{j}{j}$. 
Assuming weak driving and tracing out the resonator degree of freedom, the drive Hamiltonian can be well approximated as \cite{gambetta2011analytic}  
\begin{equation}
    \hat{H}_{\textrm{drive}}= \varepsilon(t)\sum_{j=1}^{d-1} \lambda_{j-1} \hat{\mathcal{P}}_{j-1,j}^x.
\end{equation}
We have $\lambda_0 = 1$, and the $\lambda_j$ can be treated as input parameters.
In the numerical simulations presented in the main text, for simplicity, the transmon is assumed to be in the nonlinear oscillator regime with $\lambda_j\approx \sqrt{j}$ and $\lambda_0 =1$.
Subscripts $j$ refer to the $j$th eigenenergy state and we defined the projector $\hat{\mathcal{P}}_{j,k}^x = \ketbra{j}{k}+\ketbra{k}{j}$.
For a circuit QED system, one needs to properly take into consideration couplings between the resonator mode and the particular eigenstate of the transmon to arrive at the correct $\lambda_j$'s and $\varepsilon(t)$ \cite{gambetta2011analytic}.
For simplicity, let us take 
\begin{equation}
    \varepsilon(t)=\Omega_x (t) \cos (\omega_d t +\phi_0) + \Omega_y (t) \sin (\omega_d t + \phi_0)
\end{equation}
as the external driving field with frequency $\omega_d$. 
The relative phase between the envelope and the carrier at the start of the operation is given by $\phi_0$.
In a rotating frame given by
\begin{equation}
    \hat{K}(t) = \sum_{j=1}^{d-1} \exp(-ij\omega_d t)\hat{P}_j,
\end{equation}
we arrive at
\begin{eqnarray}\label{eq:DRAG_H}
    \hat{H}'_{\textrm{gate}}&=& \hat{K}^\dagger (t) \hat{H}_{\textrm{gate}} \hat{K}(t) + i \dot{\hat{K}}^\dagger (t) \hat{K}(t)\nonumber \\
    &=& \sum_{j=1}^{d-1} (j\delta(t)+\Delta_j)\hat{P}_j \\ &+&\sum_{j=1}^{d-1} \lambda_{j-1} \left[\frac{\Omega_x (t)}{2}\hat{\mathcal{P}}_{j-1,j}^x +\frac{\Omega_y (t)}{2} \hat{\mathcal{P}}_{j-1,j}^y \right]\nonumber
\end{eqnarray}
after the rotating wave approximation.
Here, $\hat{\mathcal{P}}_{j,k}^y = -i\ketbra{j}{k}+i\ketbra{k}{j}$ and $\delta(t)=\omega_1 (t)-\omega_d$.
Ideally, the above Hamiltonian generates a unitary evolution
\begin{equation}
    \hat{\mathcal{U}}=\mathcal{T} \exp \left[-i\int_0 ^{t_g} \hat{H}'_{\textrm{gate}}(t) dt \right]=e^{i\phi_0 }\hat{\mathcal{U}}_{\textrm{qubit}}\oplus\hat{\mathcal{U}}_{\textrm{rest}}.
\end{equation}
For a leakage-free qubit \cite{motzoi2009simple,gambetta2011analytic,de2015fast}, we require $\delta(t)=\Omega_y(t)=0.$ 
The Gaussian modulated envelope \cite{gambetta2011analytic}
\begin{equation}
    \Omega_x (t)= A\left(\frac{\exp\left[\frac{-(t-t_\textrm{g}/2)^2}{2\sigma^2} \right]-\exp \left(-\frac{t_\textrm{g} ^2}{8\sigma^2} \right)}{\sqrt{2\pi\sigma^2}\textrm{erf}\left(\frac{t_\textrm{g}}{\sqrt{8}\sigma} \right)-t_\textrm{g} \exp \left(-\frac{t_\textrm{g} ^2}{8\sigma^2} \right)} \right)
\end{equation}
is chosen such that it starts and ends at zero.
The parameter $t_\textrm{g}$ refers to the gate time, while $\sigma$ is the standard deviation of the Gaussian distribution.
In the limit $t_\textrm{g} \rightarrow \infty$, it recovers the standard Gaussian distribution function.
The only requirement here is that $\int_0 ^{t_\textrm{g}}\Omega_x (t) dt=\pi$, since we want a bit-flip gate.
Hence, $A=\pi$.
Lastly, we note that \eqref{eq:DRAG_H} is the Hamiltonian we use to digitally simulate the bit-flip gate in our numerical experiments.


\section{Average gate fidelity}
\label{app:sec:gates_fidelity}
As seen in the main text and in Sec.~\ref{app:sec:drag}, due to the weak anharmonicity and multi-level nature of the transmon, there is undesired leakage during the single- and two-transmon gate evolution. 
In general, leakage is modeled by describing the system of interest as a subsystem of a larger Hilbert space governed by unitary evolution.
Hence, one can label $d_1$ energy levels where the computational subspace is $\mathcal{H}_1$, and all the other energy levels ($d_2$-dimensional) are labeled as $\mathcal{H}_2$. 
For example, a single transmon has $\mathcal{H}_1 = \text{span}\{\ket{0},\ket{1} \}$, while two transmons have $\mathcal{H}_1 = \text{span}\{\ket{00},\ket{01}, \ket{10},\ket{11} \}$. 
The $d_2$-dimensional subspace of all additional levels that the system may arrive at due to some leakage dynamics can be called the leakage subspace.
The total state space is $(d_1 + d_2)$-dimensional and denoted as $\mathcal{H} =\mathcal{H}_1 \oplus \mathcal{H}_2$.
With leakage, we would like to quantify how our unitary gate evolution performs.
This is done using the average gate fidelity \cite{bowdrey2002fidelity,wood2018quantification}, defined as
\begin{eqnarray}
    \Bar{F} &=&\int_{\psi_1 \in \mathcal{H}_1} d\psi_1 \textrm{Tr}\left(\hat{U}\ketbra{\psi_1}{\psi_1}\hat{U}^\dagger \hat{\mathcal{G}}[\ketbra{\psi_1}{\psi_1}] \right)\\
    &\approx& \frac{1}{\mathcal{M}}\sum_{i=1}^\mathcal{M} \textrm{Tr}\left(\hat{U}\ketbra{\psi_{1}^{(i)}}{\psi_{1}^{(i)}}\hat{U}^\dagger \hat{\mathcal{G}}\left[\ketbra{\psi_{1}^{(i)}}{\psi_{1}^{(i)}}\right] \right),\nonumber
\end{eqnarray}
where $\hat{U} = \hat{U}_{\textrm{des}}$ is a unitary map and $\hat{\mathcal{G}}$ is a general linear, trace-preserving quantum channel acting on an initial pure state $\ketbra{\psi_1}{\psi_1}$.
In the context of this work, $\hat{\mathcal{G}} = \hat{U}_{\textrm{ex}}$ (left hand side of \eqref{eq:Trotterized_unitary}).
The integral is over the uniform (Haar) measure $d\psi_1$ over the computational subspace $\mathcal{H}_1$, normalized to $\int d\psi_1 =1$.
A fidelity of $\Bar{F}=1$ would mean that $\hat{U}$ implements $\hat{\mathcal{G}}$ perfectly.
\\
\\

\bibliographystyle{apsrev4-1}
\bibliography{main}

\end{document}